# Osmium-nitrosyl oxalato-bridged lanthanide-centred pentanuclear complexes: Synthesis, crystal structures and magnetic properties

Anatolie Gavriluta,[a,b] Nicolas Claiser,[c] Paul-Steffen Kuhn,[b] Ghenadie Novitchi,[d] Jean Bernard Tommasino,[a] Olga Iasco,[a] Vadim Druta,[a,e] Vladimir B. Arion*[,b] and Dominique Luneau*[,a]

[a] *Université Claude Bernard Lyon 1, Laboratoire des Multimatériaux et Interfaces (UMR 5615), Campus de la Doua, 69622 Villeurbanne cedex, France*

[b] *University of Vienna, Faculty of Chemistry, Institute of Inorganic Chemistry, Währinger Strasse 42, 1090 Vienna, Austria*

[c] *Université de Lorraine, Laboratoire de Cristallographie, Résonance Magnétique et Modélisations, UMR CNRS 7036, Vandœuvre-lès-Nancy, 54506 France*

[d] *Laboratoire National des Champs Magnétiques Intenses-CNRS, Université Joseph Fourier, 25 Avenue des Martyrs, 38042 Grenoble Cedex 9, France*

[e] *Institute of Chemistry of the Academy of Sciences of Moldova, str. Academiei 3, MD 2028, Chisinau, Moldova*

Corresponding Authors :
*E-mail: vladimir.arion@univie.ac.at (V. B. A.); dominique.luneau@univ-lyon1.fr (D. L.)

**Abstract:** A series of pentanuclear heterometallic coordination compounds of the general formula $(Bu_4N)_5[Ln\{Os(NO)(\mu\text{-}ox)Cl_3\}_4(H_2O)_n]$, where Ln = Y (**2**) and Dy (**3'**), when n = 0, and Ln = Dy (**3**), Tb (**4**), and Gd (**5**), when n = 1, has been synthesized from the reaction of the precursor $(Bu_4N)_2[Os(NO)(ox)Cl_3]$ (**1**) with the respective lanthanide(III) (Gd, Tb, Dy) or yttrium(III) chloride. The coordination numbers eight or nine are found for the central metal ion within the five new complexes. The compounds were fully characterized by elemental analysis, IR spectroscopy, single crystal X-ray diffraction, magnetic susceptibility, and ESI mass spectrometry. In addition, compound **1** was studied by UV-vis spectroscopy and cyclic voltammetry. The X-ray diffraction crystal structures have revealed that anionic complexes consist of a lanthanide or yttrium core, bridged via oxalate ligands with four octahedral osmium-nitrosyl moieties. This picture, in which the central ion is eight-coordinate, holds for the lanthanide ions with an ionic radius smaller than that of the dysprosium(III) ion. For larger ionic radii, the central metal ion is nine-coordinate as the coordination sphere is completed by one molecule of water. Only in the case of dysprosium(III) it was possible to obtain complexes with both coordination numbers 8 and 9 thus implying that dysprosium(III) is the tilt limit between the two coordination numbers in this series. The bond length Ln–OH$_2$ decreases from Dy to Gd. The nine-coordinate complexes are energetically more favored for lanthanide ions with a radius larger than that of the dysprosium(III) and the eight-coordinate for smaller ions. The magnetic studies of the series of compounds have shown that the osmium precursor **1** as well as the yttrium compound **2** are diamagnetic, while the magnetism of the gadolinium, terbium and dysprosium complexes is due to isolated lanthanide ions.

# The Table of Contents

**Graphic (8 cm large 4 cm high)**

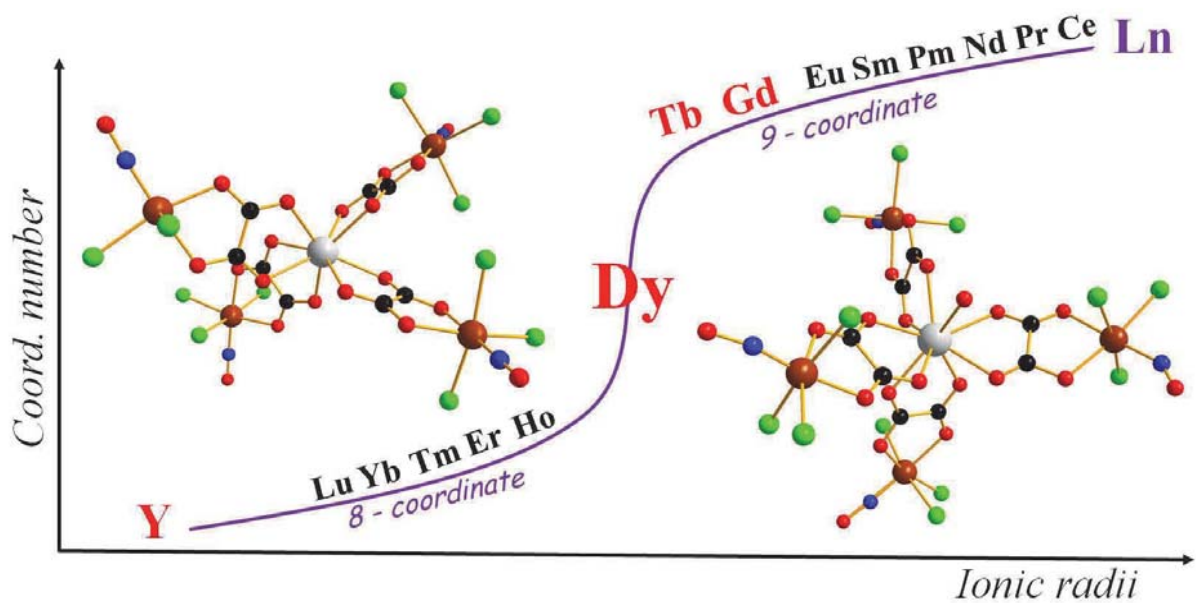

**Text (20-30 word):**

**Coordination tilt:** osmium-nitrosyl oxalato-bridged lanthanide-centred pentanuclear complexes exhibit the two coordination numbers 8 to 9 for the central lanthanide that tilts on dysprosium.

# Introduction

Polynuclear metal complexes are of interest in many areas of research, ranging from models of biological systems, such as the photosynthetic water oxidation site,[1] to materials comprising photoactive systems,[2] molecular electronics,[3] molecular magnetism,[4] and single-molecule magnets,[5, 6, 7] as well as in the development of anticancer drugs.[8] From a coordination chemist's point of view, most of these fields share some common features. Overall, the synthesis approach determines the route for mastering the type, number and positions of the different metal ions to be assembled. In light of this the field of molecular magnetism, and more specifically that of single-molecule magnets, has been dramatically imaginative and productive in the design and elaboration under some extent of control of coordination clusters with all sorts of nuclearity and metal combination.[5, 9]

The predictability of the polynuclear architecture mainly depends on the choice of the ligands and despite of some success[10] it is never an easy task. In that regard, among numerous ligands, oxalate[11] and its thio-[12] and oxamide-derivatives[13] have proved to be very efficient in building polynuclear assemblies in a quite reliable way. Having this in mind, as part of our interest in both, magnetism of polynuclear metal complexes[10a, 14] and osmium and ruthenium-nitrosyl complexes as potential anticancer drugs,[15] we investigated the ability of the osmium-oxalato and osmium-nitrosyl-oxalato complexes to form polynuclear systems with different lanthanides.

As for molecular magnetism and single-molecule magnets, the 4-5d and 4-5f metal ions are very attractive. Indeed, their strong and well-known spin-orbit coupling[4a, 16] is expected to favor large magnetic anisotropy and high blocking temperatures for the reversal of magnetization. This has been demonstrated for rare earth metal ions, where single ion complexes of lanthanides have shown SMM behavior.[17] As a result, the number of polynuclear systems of lanthanide ions, single or combined with 3d metal ions, has increased dramatically over the last years[6b, 14, 18] and many of them exhibit a large energy barrier for the reversal of the magnetization.[19] In addition, some actinide complexes have also been found to possess SMM behavior.[20] In contrast, SMM based on 4d and 5d elements, alone[21] or with 3d[22] and 4f[23] metal ions, are still far less numerous. This may be due to the non-trivial synthesis approaches that are usual for the 4d and 5d metal ions. However, despite these difficulties, it seems to us important to be involved in this area that may be rich in magnetic behaviors.[16, 24]

In the wake of platinum, 4d and 5d metal ions are also attractive for development of anticancer drugs.[25] Therefore, our research interests focused on osmium and ruthenium-nitrosyl complexes, and, in particular, on complexes with azole heterocycles, or amino acids as ancillary ligands.[15] In addition, we investigated oxalato complexes having in mind that the platinum-oxalato derivative, namely, oxaliplatin, is an efficient antitumor agent,[25-26] and the ability of oxalate to act as a good bridging ligand. This last feature can be explored for the synthesis of heterometallic systems with enhanced anticancer properties and interesting magnetic behavior.[11] Moreover, we are interested to develop a drug whose antiproliferative activity may be solely triggered once inside the cancerous cells, for example upon photoactivation, to release high concentrations of free nitric oxide in combination with an activated anticancer metal complex.

Herein we report on the synthesis, crystal structures, magnetic and electrochemical properties of a series of osmium-nitrosyl oxalato-bridged lanthanide-centred pentanuclear complexes with eight-coordination of the lanthanide ion $(Bu_4N)_5[Ln\{Os(NO)(\mu\text{-}ox)Cl_3\}_4]$ for Ln = Y (**2**) and Dy (**3'**) and with nine-coordination $(Bu_4N)_5[Ln\{Os(NO)(\mu\text{-}ox)Cl_3\}_4(H_2O)]$ for Ln = Dy (**3**), Tb (**4**) and Gd (**5**) (**Chart 1**), where ox stands for oxalate and Bu for *n*-butyl. All complexes were obtained by reacting the precursor $(Bu_4N)_2[Os(NO)(ox)Cl_3]$ (**1**) with the respective lanthanide(III) (Gd, Tb, Dy) or yttrium(III) chloride salt. Similar complexes of the formula $(Bu_4N)_5[Gd\{ReBr_4(\mu\text{-}ox)\}_4(H_2O)]$[27] as well as the related $(NBu_4)_4[Ni\{ReCl_4(\mu\text{-}ox)\}_3]$[28] were obtained using $(NBu_4)_2[ReX_4(ox)]$ (X= Cl, Br)[13a] as starting material.

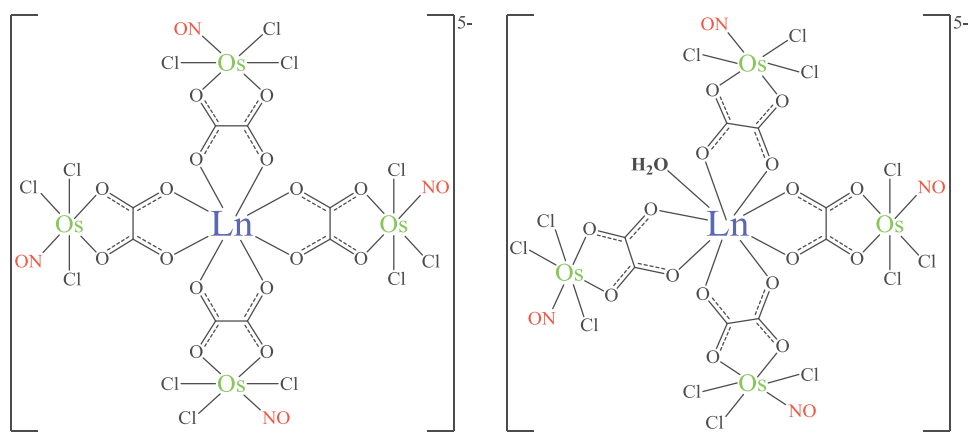

**Chart 1.** Polynuclear lanthanide-Os(NO) complexes with the lanthanide coordination numbers 8 and 9.

## Experimental Section

**Synthesis:** OsO$_4$ (99.8%) was purchased from Johnson Matthey, while YCl$_3$·6H$_2$O, GdCl$_3$·6H$_2$O, DyCl$_3$·6H$_2$O, TbCl$_3$·6H$_2$O were obtained from Strem Chemicals and NH$_2$OH·HCl, K$_2$C$_2$O$_4$·H$_2$O, H$_2$C$_2$O$_4$·2H$_2$O and NH$_4$PF$_6$, Bu$_4$NCl from Acros and Sigma Aldrich, respectively. The starting compound (Ph$_4$P)$_2$[Os(NO)Cl$_5$] was synthesized as previously reported in the literature.[29]

**(Bu$_4$N)$_2$[Os(NO)Cl$_3$(ox)] (1).** A solution of NH$_4$PF$_6$ (0.98 g, 6.0 mmol) in methanol (20 mL) was added to a solution of (Ph$_4$P)$_2$[Os(NO)Cl$_5$][29] (3.24 g, 3.0 mmol) in methanol (200 mL) under stirring at room temperature for 10 min and the resulting precipitate (Ph$_4$P)PF$_6$, was filtered off from the reaction mixture. The filtrate was evaporated to dryness to give (NH$_4$)$_2$[Os(NO)Cl$_5$] and an excess of oxalic acid (5.0 g, 86 mmol) and water (50 mL) was added. The resulting solution was refluxed for 6 h and, after cooling to room temperature, a solution of Bu$_4$NCl (2.0 g, 7.2 mmol) in water (30 mL) was slowly added. The resulting red precipitate of **1** was filtered off and washed with water (3 x 5 mL), ethanol / water (1:1) (3 x 5 mL), and diethyl ether (3 x 3 mL), and dried *in vacuo*. Yield: 2.02 g, 72 %. Analytical data for **1**: Anal. Calcd for C$_{34}$H$_{72}$N$_3$O$_5$Cl$_3$Os ($M$ = 899.54 g/mol): C, 45.40; H, 8.07; N, 4.67. Found: C, 45.44; H, 7.76; N, 4.61. ESI-MS in CH$_3$CN (negative): *m/z* 657 [Os(NO)(ox)Cl$_3$+Bu$_4$N]$^-$, 380 [Os(NO)(ox)Cl$_2$]$^-$, 362 [Os(NO)Cl$_3$+Cl]$^-$, 327 [Os(NO)Cl$_3$]$^-$. IR, cm$^{-1}$: 460, 548, 627, 742, 802, 882, 1030, 1068, 1151, 1221, 1352, 1481, 1666, 1697, 1789, 2874, and 2959. UV–vis in CH$_3$CN, λ$_{max}$, nm (ε, M$^{-1}$cm$^{-1}$): 230 (12620), 276 (1738), 303 (1306), 416 (157), 457 (150), 547 (145). Suitable crystals of **1** for X-ray diffraction study were obtained by re-crystallization from acetonitrile.

**(Bu$_4$N)$_5$[Ln{Os(NO)(μ-ox)Cl$_3$}$_4$(H$_2$O)$_n$]**, where **Ln = Y (2)**, **Dy (3')**, when **n = 0**, and **Ln = Dy (3)**, **Tb (4)**, **Gd (5)**, when **n = 1**. A solution of LnCl$_3$·6H$_2$O (0.03 mmol) in 2-propanol (1.5 mL) was added to **1** (90 mg, 0.1 mmol) in acetonitrile (2.5 mL) and the mixture was refluxed for 40 min. Well-shaped red hexagons were formed by slow evaporation of the solvent during 3–5 days. The crystalline product was filtered off, washed with diethyl ether and dried *in vacuo*. Yield of **2**: 44 mg, 59%. Analytical data for **2**: Anal. Calcd for C$_{88}$H$_{180}$N$_9$O$_{20}$Cl$_{12}$Os$_4$Y ($M$ = 2959.68 g/mol): C, 35.71; H, 6.13; N, 4.26. Found: C, 35.44; H, 5.87; N, 4.61. IR, cm$^{-1}$: 412, 458, 544, 630, 737, 813, 883, 1027, 1060, 1107, 1153, 1286, 1334, 1379, 1476, 1626, 1651, 1817, 2874, and 2960.

Yield of **3'**: Only a few crystals were isolated using dry solvents.

Yield of **3**: 50 mg, 65%. Analytical data for **3**: Anal. Calcd for $C_{88}H_{182}N_9O_{21}Cl_{12}Os_4Dy$ ($M$ = 3051.29 g/mol): C, 34.64; H, 6.01; N, 4.13. Found: C, 34.74; H, 5.87; N, 4.10. IR, cm$^{-1}$: 413, 458, 543, 630, 737, 812, 883, 1027, 1064, 1107, 1153, 1287, 1333, 1379, 1477, 1624, 1647, 1817, 2874, 2960, and 3342.

Yield of **4**: 57 mg, 75%. Analytical data for **4**: Anal. Calcd for $C_{88}H_{182}N_9O_{21}Cl_{12}Os_4Tb$ ($M$ = 3047.72 g/mol): C, 34.68; H, 6.02; N, 4.14. Found: C, 34.84; H, 5.96; N, 4.03. IR, cm$^{-1}$: 410, 458, 544, 630, 737, 812, 882, 1027, 1065, 1107, 1153, 1285, 1327, 1379, 1445, 1462, 1633, 1658, 1816, 2874, 2960, and 3339.

Yield of **5**: 53 mg, 69%. Analytical data for **5**: Anal. Calcd for $C_{88}H_{182}N_9O_{21}Cl_{12}Os_4Gd$ ($M$ = 3046.04 g/mol): C, 34.70; H, 6.02; N, 4.14. Found: C, 34.87; H, 5.93; N, 4.01. IR, cm$^{-1}$: 407, 458, 544, 737, 811, 881, 1028, 1063, 1107, 1152, 1284, 1325, 1380, 1443, 1464, 1634, 1655, 1814, 2873, 2960, and 3340.

Single crystals of **2, 3', 3, 4,** and **5** suitable for X-ray diffraction studies were obtained from the corresponding reaction mixture.

***Physical Measurements.*** Elemental analyses were performed by the Microanalytical Service of the Faculty of Chemistry of the University of Vienna. IR spectra were recorded in the solid state on a NICOLET spectrophotometer in the 400−4000 cm$^{-1}$ range, while the UV–vis spectrum was recorded on a Perkin-Elmer Lambda 35 using CH$_3$CN as solvent. Mass spectra were recorded on an ion trap mass spectrometer (LCQ, Thermo, Bremen, Germany) equipped with an electrospray ion source (ESI) in the positive and negative ion mode. The spray voltage for the positive and negative ion mode is 4 kV and -3 kV, respectively, using a capillary transfer temperature of 200 °C.

***Electrochemical Measurements*** were performed using an AMEL 7050 all-in one potentiostat, equipped with a standard three-electrode set up containing a glassy carbon electrode, a platinum auxiliary electrode and a saturated calomel electrode (SCE) as reference electrode. Degassing of solutions was accomplished by passing a stream of N$_2$ through the solution for 30 min prior to the measurement and then maintaining a blanket atmosphere of N$_2$ over the solution during the measurement. The potentials were measured in a freshly prepared complex solution in CH$_3$CN (1 or 2 mM) containing 0.1 M (*n*-Bu$_4$N)PF$_6$ as supporting electrolyte. Under these experimental conditions, the ferrocene / ferrocenium couple, used as an internal reference for potential measurements, was located at E$_{1/2}$ = +0.425 V.

*Magnetic susceptibility* data (2-300 K) were collected on powdered samples using a SQUID magnetometer (Quantum Design MPMS-XL), applying a magnetic field of 0.1 T. Magnetization isotherms were collected at 2.0 K between 0 and 5 T. All data were corrected for the contribution of the sample holder and the diamagnetism of the samples was estimated from Pascal's constants.[4a, 30]

*Crystallographic Structure Determination*. All data collections were conducted at CRM2 laboratory. Due to poor crystal quality, several samples of each complex were tested. Crystals obtained in Lyon were tested at room temperature on a Bruker kappa CCD diffractometer equipped with an APEX2 detector and the best samples were measured at 100 K for a complete data collection on an Agilent SuperNova diffractometer equipped with an ATLAS detector. The CrysAlisPro suite was used to determine the unit cell and the data collection strategy.[31] The data treatment and reduction including absoption correction was carried out using CrysAlisPro suite. Structure solution and refinement was conducted with the WinGX package.[32] All atomic displacements parameters for non-hydrogen atoms have been refined with anisotropic terms except some atoms presenting high disorder, especially carbon atoms of the $Bu_4N$ (isotropic terms). Some of the hydrogen atoms were located by difference Fourier maps, the rest was theoretically located on the basis of the conformation of the supporting atom and refined keeping restraints (*riding mode*).

The crystal data and refinement parameters for compounds **1**, **2, 3, 3', 4,** and **5** are summarized in Table 1. Selected bond lengths and angles are given in Tables 2–4.

## Results

**Synthesis**

Complex **1** was prepared under reflux in aqueous media for 6 h by the reaction between oxalic acid and $(NH_4)_2[Os(NO)Cl_5]$. The latter one was previously obtained by stoichiometric reaction of $(Ph_4P)_2[Os(NO)Cl_5]$ and $NH_4PF_6$ in methanol with almost complete conversion after precipitation of $(Ph_4P)PF_6$. The complex $[Os(NO)(ox)Cl_3]^{2-}$ was precipitated as the tetrabutylammonium salt to give **1** with an overall yield of 75 %.

Complexes **2**–**5** were prepared by the reaction of **1** with a small excess of of the respective chloride salt of Dy(III), Tb(III), Gd(III) and Y(III) in a mixture of acetonitrile and 2-propanol at reflux for 40 min with an average yield of 60 – 75%.

Complexes of Tb(III) (**3**) and Gd(III) (**5**) were obtained only with nine-coordination and all attempts to obtain eight-coordinate complexes using dry solvents were unsuccessful. Y(III) gave only the eight-coordinate complex **2**. Dy(III) gave easily the nine-coordinate complex **3**. However, it was also possible to obtain some single crystals of the eight-coordinate complex **3'** using dry solvents, but they were not stable in time and only the X-ray crystal structure could be determined. The peak with *m/z* 657 in the negative mode ESI mass spectra of **1** was assigned to [Os(NO)(ox)Cl$_3$+Bu$_4$N]$^-$, while signals at *m/z* 380, 362 and 327 were attributed to [Os(NO)(ox)Cl$_2$]$^-$, [Os(NO)Cl$_3$+Cl]$^-$ and [Os(NO)Cl$_3$]$^-$, respectively. The presence of the NO moiety was confirmed by IR spectroscopy, where the characteristic NO band was found at 1788 cm$^{-1}$ for precursor **1,** while a hypsochromic shift to 1814 – 1817 cm$^{-1}$ was observed for complexes **2 – 5**. The carboxylic groups were also detected in the IR spectra, where for **1**, $\nu_{as}$(COO$^-$) was seen as two peaks at 1666 and 1697 cm$^{-1}$ and $\nu_s$(COO$^-$) at 1352 cm$^{-1}$, but in case of the **2 – 5**, $\nu_{as}$(COO$^-$) appeared as two peaks between 1624 – 1658 cm$^{-1}$ and $\nu_s$(COO$^-$) in the range 1326 – 1334 cm$^{-1}$ and 1378 – 1379 cm$^{-1}$, respectively.

The syntheses of the analogous complexes without nitrosyl (NO), using (Bu$_4$N)$_2$[OsCl$_4$(ox)] as precursor, were unsuccessful despite many tentatives. This is in contrast to the previously reported rhenium congener.[27]

**Table 1**: Crystallographic data and refinement details for compounds **1**–**5**.

| | **1** (Os) | **2** (Y-Os) | **3'** (Dy-Os) | **3** (Dy-Os) | **4** (Tb-Os) | **5** (Gd-Os) |
|---|---|---|---|---|---|---|
| Formula | $C_{34}H_{72}N_3O_5Cl_3Os$ | $C_{88}H_{180}N_9O_{20}Cl_{12}Os_4Y$ | $C_{88}H_{180}N_9O_{20}Cl_{12}Os_4Dy$ | $C_{88}H_{182}N_9O_{21}Cl_{12}Os_4Dy$ | $C_{88}H_{182}N_9O_{21}Cl_{12}Os_4Tb$ | $C_{88}H_{182}N_9O_{21}Cl_{12}Os_4Gd$ |
| $M$ [g/mol] | 899.52 | 2959.58 | 3033.17 | 3051.19 | 3047.61 | 3045.94 |
| $\lambda$ [Å] | 0.71073 | 0.71073 | 0.71073 | 0.71073 | 0.71073 | 0.71073 |
| $T$ [K] | 100(2) | 100(2) | 100(2) | 100(2) | 100(2) | 100(2) |
| Space group. | $P2_1/c$ | $P\text{-}42_1$ | $P\text{-}42_1$ | $Cc$ | $Cc$ | $Cc$ |
| $a$ [Å] | 19.5248(2) | 18.437(5) | 18.4332(1) | 17.9951(2) | 18.0677(4) | 17.9574(3) |
| $b$ [Å] | 14.7764(1) | 18.437(5) | 18.4332(1) | 27.0467(3) | 27.0766(6) | 26.8448(5) |
| $c$ [Å] | 30.5673(2) | 18.446(5) | 18.4529(1) | 26.8732(4) | 26.9201(9) | 26.7365(5) |
| $\alpha$ [°] | 90 | 90 | 90 | 90 | 90 | 90 |
| $\beta$ [°] | 106.870(1) | 90 | 90 | 92.681(1) | 92.351(3) | 92.519(2) |
| $\gamma$ (°) | 90 | 90 | 90 | 90 | 90 | 90 |
| $V$ [Å$^3$] | 8439.34(12) | 6270(3) | 6269.98(6) | 13065.1(3) | 13158.5(6) | 12876.2(4) |
| $Z$ | 8 | 8 | 8 | 4 | 4 | 4 |
| $D$ [g/cm$^3$] | 1.416 | 1.568 | 1.601 | 1.551 | 1.538 | 1.571 |
| $\mu$ [mm$^{-1}$] | 3.251 | 4.810 | 4.940 | 4.740 | 4.679 | 4.747 |
| $\Theta$ range [°] | 2.61 - 45.29 | 2.47 - 45.39 | 2.47 - 33.14 | 2.62 - 41.00 | 2.52 - 32.59 | 2.54 - 30.51 |
| Reflns Coll. | 438132 | 196891 | 248299 | 297405 | 119805 | 133760 |
| Ind. Refl. | 70611 | 9598 | 11956 | 82469 | 41110 | 39132 |
| Ind. Refl. [I > 2sig(I)] | 52820 | 8998 | 11595 | 36259 | 28316 | 30958 |
| Params | 1405 | 363 | 446 | 935 | 813 | 1073 |
| GOF on F$^2$ | 1.017 | 1.47 | 1.194 | 1.250 | 1.036 | 1.620 |
| $R1^{(a)}$ [I > 2sig(I)] | 0.0386 | 0.0758 | 0.0242 | 0.1351 | 0.1007 | 0.1339 |
| $wR2^{(b)}$ (all) | 0.0782 | 0.1809 | 0.0620 | 0.4269 | 0.2668 | 0.4073 |

$^{(a)}R_1 = \Sigma||Fo|-|Fc||/\Sigma|Fo|$, $^{(b)}wR_2 = [\Sigma(w(Fo^2-Fc^2)^2)/\Sigma(w(Fo^2)^2)]^{1/2}$ with $w = 1/[(\sigma^2Fo^2)+(aP)^2+bP]$ and $P = (\max(Fo^2)+2Fc^2)/3$

**Crystal structure.** The crystal data and refinement parameters for compounds **1**, **2, 3', 3, 4,** and **5** are summarized in Table 1.

**(Bu₄N)₂[Os(NO)Cl₃(ox)] (1).** The precursor crystallizes in the monoclinic space group $P2_1/c$ with two osmium complex anions [Os(NO)Cl₃(ox)]²⁻ and four tetrabutylammonium cations (Bu₄N⁺) in the asymmetric unit. Each osmium complex is surrounded by six tetrabutylammonium cations forming a sphere of approximately 8.5 Å radius, in which the shortest Os-Os distance is 9.833 Å. The bond lengths corresponding to the Os coordination sphere are quoted in Table 2 with atom labelling as shown in Figure 1. For more details, the corresponding CIF is given in supplementary materials. The Os-Cl bond lengths range between 2.3481(9) and 2.3798(6) Å and the Os-O bond lengths are between 2.019(2) and 2.056(2) Å.

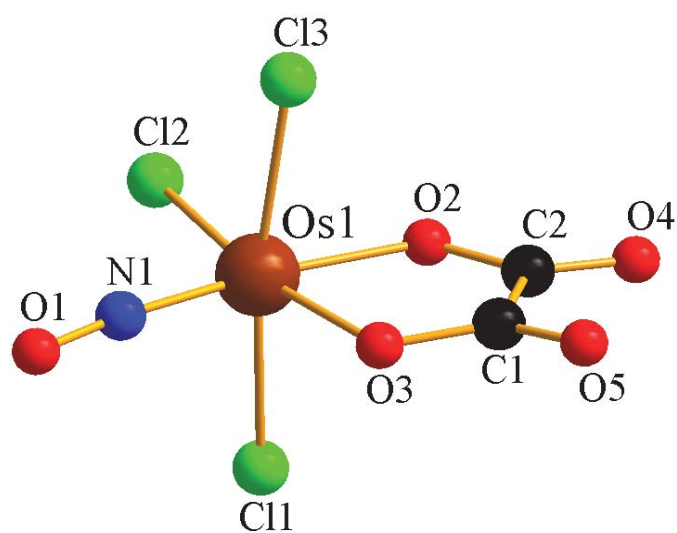

**Figure 1:** View of complex **1**. Hydrogen atoms and tetrabutylammonium cations (Bu₄N⁺) are omitted for clarity. Only numbering of the atom labels is shown as the same numbering was used for two crystallographically independent molecules.

**Table 2**: Selected bond lengths (Å) for compound **1**.

|         | Molecule A | Molecule B |
|---------|------------|------------|
| Os1-Cl1 | 2.3784(3)  | 2.3663(4)  |
| Os1-Cl2 | 2.3481(6)  | 2.3633(4)  |
| Os1-Cl3 | 2.3791(4)  | 2.3803(4)  |
| Os1-O2  | 2.023(1)   | 2.020(1)   |
| Os1-O3  | 2.043(1)   | 2.057(1)   |
| Os1-N1  | 1.805(2)   | 1.742(2)   |
| O1-N1   | 1.031(4)   | 1.031(4)   |

**(Bu$_4$N)$_5$[Ln{Os(NO)(μ-ox)Cl$_3$}$_4$] where Ln = Y (2), Dy (3').** These 8-coordinate lanthanide complexes crystallize in the tetragonal space group P-42$_1$c. The central ion (Y or Dy) and one nitrogen atom of a tetrabutylammonium cation lie on the P-4 axis along the *c* axis. The four oxalato ligands of the central lanthanide cation are thus symmetrically related as well as the alkyl chains of the tetrabutylammonium cation. A second tetrabutylammonium ion is in general position leading to a total of five tetrabutylammonium cations for each osmium-lanthanide complex. Note that the atoms of the tetrabutylammonium cation in a special position show higher thermal smearing, certainly linked to high static disorder. In these complexes, each Os-oxalate moiety is surrounded by 5 tetrabutylammonium cations forming, as in the precursor (**1**), a sphere of approximately 8.5 Å radius. In case of the Y complex the shortest Os-Os distance is of 7.496(2) Å and 7.496(2) Å for the Dy crystal. The Y-Y and Dy-Dy distances are 15.970 and 15.769 Å, respectively. The bond lengths corresponding to the heavy atom coordination spheres are given in Table 3. For more details, the corresponding CIFs are given in supplementary materials.

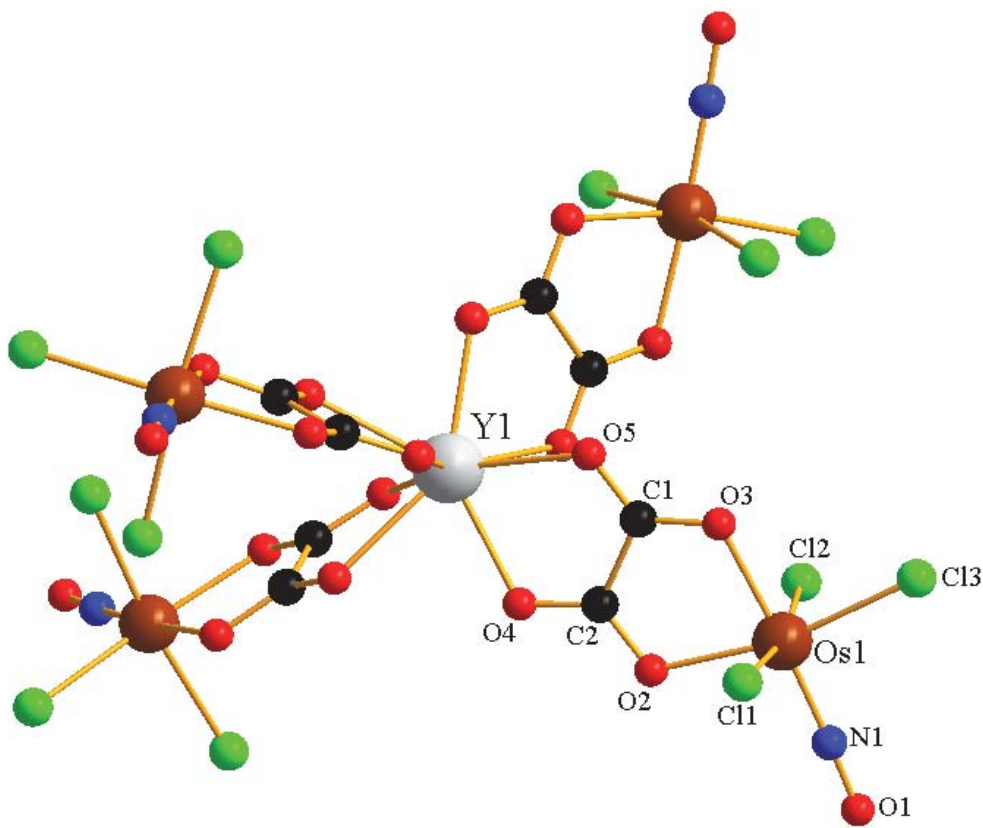

**Figure 2:** View of complex **2**. Tetrabutylammonium cations (Bu$_4$N$^+$) and hydrogen atoms are omitted for clarity.

**Table 3**: Selected bond lengths (Å) for compounds **2** and **3'**.

|         | **2** (Y)  | **3'** (Dy) |
|---------|-----------|-------------|
| Os1-Cl1 | 2.371(2)  | 2.3713(7)   |
| Os1-Cl2 | 2.330(2)  | 2.3295(8)   |
| Os1-Cl3 | 2.346(3)  | 2.3483(9)   |
| Os1-O2  | 2.073(6)  | 2.059(2)    |
| Os1-O3  | 2.085(6)  | 2.084(2)    |
| Os1-N1  | 1.743(8)  | 1.745(3)    |
| O1-N1   | 1.15(1)   | 1.154(4)    |
| Ln1-O4  | 2.394(6)  | 2.400(2)    |
| Ln1-O5  | 2.327(6)  | 2.344(2)    |

Considering the values given in Table 3, it appears that the Dy cation radius is approximately by 0.01 Å larger than the Y one (at 100 K). The Os-Cl bonds range from 2.3478(9) to 2.372(1) Å and are in agreement with data observed for **1**. The Os-O bond lengths are approximately by 0.04 Å longer than those found in **1**. These differences are ascribed to the different coordination mode of oxalate ligand in mononuclear species **1** and in heteronuclear complexes **2** and **3**. In the first one the $C_2O_4^{2-}$ acts as bidentate, while in **2** and **3** as a bridging tetradentate ligand between osmium and lanthanide ion.

**$(Bu_4N)_5\{[Os(NO)(\mu\text{-}ox)Cl_3]_4Ln(H_2O)\}$** wher**e Ln = Dy (3), Tb (4), Gd (5) :** These 9-coordinate complexes crystallize in the monoclinic space group Cc. For these coordination compounds, the crystal quality was poor and they do not support the lowering in temperature. As a result both Gd and Tb structures are incomplete at 100 K. In the case of the Tb-centered compound **4**, five carbon atoms of the ligands could not be precisely positioned due to a large disorder. Nevertheless, it appears that the central ion is surrounded by 9 oxygen atoms originating from 4 oxalato ligands and one water molecule. The Gd structure is worse defined, even for the osmium coordination sphere. In this case, the Gd structure obtained at room temperature is better than at 100 K due to the degradation of the crystal quality with decreasing temperature.

Concerning the packing, the unit cell can be roughly divided in 8 parts: 4 lanthanide-osmium complexes occupied the summits of a tetrahedron included in this unit cell and the 4 other corners of the cell are occupied by the tetrabutylammonium cations. The lanthanide atoms are then well separated from each other: at 100 K, for Gd, the shorter Gd-Gd distance is 15.775 Å, in case of Tb 15.903 Å. The average Tb-O distance is 2.418 Å (standard deviation: 0.027 Å) and close to 2.43 Å for the Gd-O bonds, indicating a slightly larger radius for the later cation (Table 4).

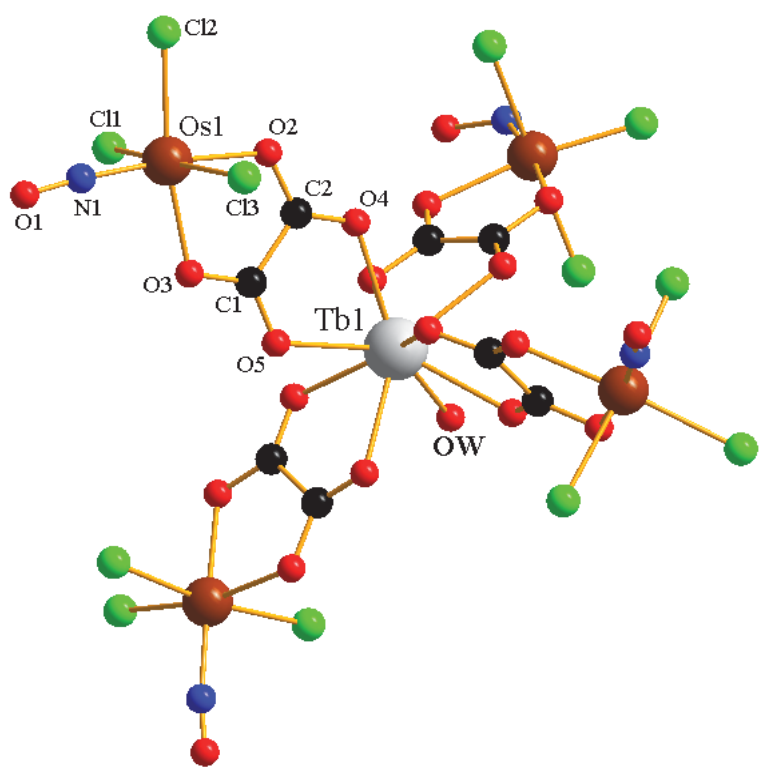

**Figure 3:** View of complex **4**. Tetrabutylammonium cations (Bu$_4$N$^+$) and hydrogen atoms are omitted for clarity. Only numbering of the atom label is shown irrespective of the A, B, C and D moieties, as the same numbering holding only for atom location was used in every moiety.

**Table 4**: Mean values for selected bond lengths (Å) for compounds **3–5**.

|         | **3** (Dy) | **4** (Tb) | **5** (Gd) |
|---------|-----------|-----------|-----------|
| Os1-Cl1 | 2.30(12)  | 2.35(4)   | 2.35(7)   |
| Os1-Cl2 | 2.23(22)  | 2.26(16)  | 2.26(13)  |
| Os1-Cl3 | 2.33(24)  | 2.33(3)   | 2.31(11)  |
| Os1-N1  | 1.67(32)  | 1.81(11)  | 1.59(35)  |
| Os1-O1  | 2.07(9)   | 2.08(6)   | 2.07(4)   |
| N1-O1   | 1.30(24)  | 1.19(19)  | 1.19(12)  |
| Ln1-O4  | 2.394(71) | 2.420(34) | 2.428(31) |
| Ln1-O5  | 2.394(14) | 2.422(21) | 2.430(30) |

**Electrochemical Measurements**

Electrochemistry studies were performed only for complex **1**. The cyclic voltammogram of **1** in CH$_3$CN containing 0.10 M TBAPF$_6$ as supporting electrolyte, using a glassy carbon working electrode and a saturated calomel electrode (SCE) as a reference electrode, is shown in Figure 4. The redox processes shown in Table 5 occur exclusively on the complex anion [Os(NO)Cl$_3$(ox)]$^{2-}$ and CVs data are difficult to interpret due to parallel electrochemical reaction on the metal centre and oxalate ligand.

The CV displays an irreversible one-electron reduction wave attributed to the {Os(NO)}$^6$ → {Os(NO)}$^7$ process with a redox potential value Ep = -1.86 V, as previously reported for osmium nitrosyl compounds,[15] followed by oxalate reduction (Ep = -2.14 V). Upon oxidation, the cyclic voltammogram shows a quasi-reversible wave at E$_{1/2}$ = 1.22 V/SCE, which consists of a one-electron oxidation reaction according to the reaction {Os(NO)}$^6$ → {Os(NO)}$^5$.

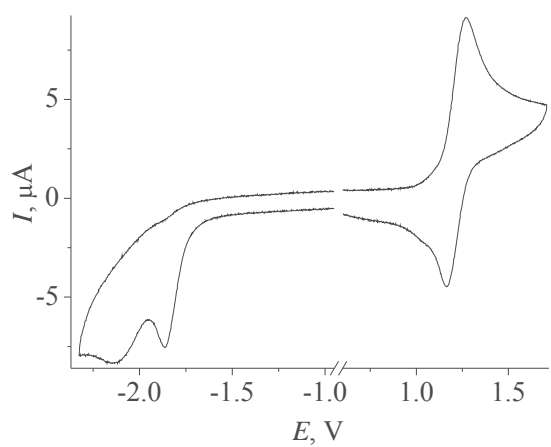

**Figure 4**. Cyclic voltammogram of **1** at 0.1 V/s on GC electrode (3 mm) in 0.1 M TBAPF$_6$ in CH$_3$CN.

**Table 5**. Electrochemical data for **1**$^a$

| Complex | cathodic peak potential ($E_{pc}$) | anodic peak potential ($E_{1/2}$) |
|---|---|---|
| 1 | −1.86$^b$ | 1.22$^c$ |

$^a$Peak potential (V) and half-wave potential $E_{1/2}$ recorded in CH$_3$CN at 293 K with a glassy carbon electrode, 0.1 M TBAPF$_6$ as supporting electrolyte; all potentials are vs SCE, scan rate 0.1 V/s; $^b$Irreversible system; $^c$Quasireversible system. Ferrocene/ferrocenium ($E_{1/2}$ = + 0.425 V) couple was used as an internal standard.

**Magnetic properties**

Magnetic properties of the homometallic osmium complex (Bu$_4$N)$_2$[Os(NO)(ox)Cl$_3$] (**1**) and the heterometallic complexes (Bu$_4$N)$_5$[Ln{Os(NO)(μ-ox)Cl$_3$}$_4$(H$_2$O)$_n$], where Ln = Y (**2**), when n = 0, and Ln = Dy (**3**), Tb (**4**), Gd (**5**), when n = 1, have been studied in the temperature range between 300–2 K with a 0.1 T dc magnetic field and in the range of the magnetic field between 0–5 T at a constant temperature of 2 K.

Compounds **1** (Os) and **2** (Os-Y) are diamagnetic in the whole temperature range. Their diamagnetism is consistent with our previous work on similar Os(NO) complexes.[15a,b] The heterometallic complexes **3**–**5** are strongly paramagnetic. The temperature dependence of the $\chi_M T$ product for **3** (Dy), **4** (Tb) and **5** (Gd) under an applied magnetic field of 0.1 T is shown in Figure 5. The $\chi_M T$ product at 300 K is 7.77, 11.41 and 14.34 cm$^3$ K mol$^{-1}$ for **3**–**5**, respectively, which is in good agreement with theoretical values expected for single isolated non-interacting lanthanide ions.[33] Upon cooling, the $\chi_M T$ product decreases in the case of **4** (Tb) and **3** (Dy) to reach the values $\chi_M T$ = 7.56 for **4** (Tb) and 9.15 cm$^3$ K mol$^{-1}$ for **3** (Dy). This corresponds to the presence of strong anisotropic components in magnetic susceptibility of dysprosium and terbium. In the case of the isotropic gadolinium analogue **5**, the $\chi_M T$ product remains constant within the limit of error of the measurements. These results suggest the absence of magnetic interaction and correspond to an isolated behavior of lanthanide ions in this heterometallic series. In addition, the magnetization measurements (0-5T) are consistent with this conclusion (Figure 5 bottom). The absence of frequency dependence of slow magnetic relaxation was proved by ac susceptibility measurements at zero and 0.2 T dc fields for the strongly anisotropic Dy and Tb compounds **4** and **5**.

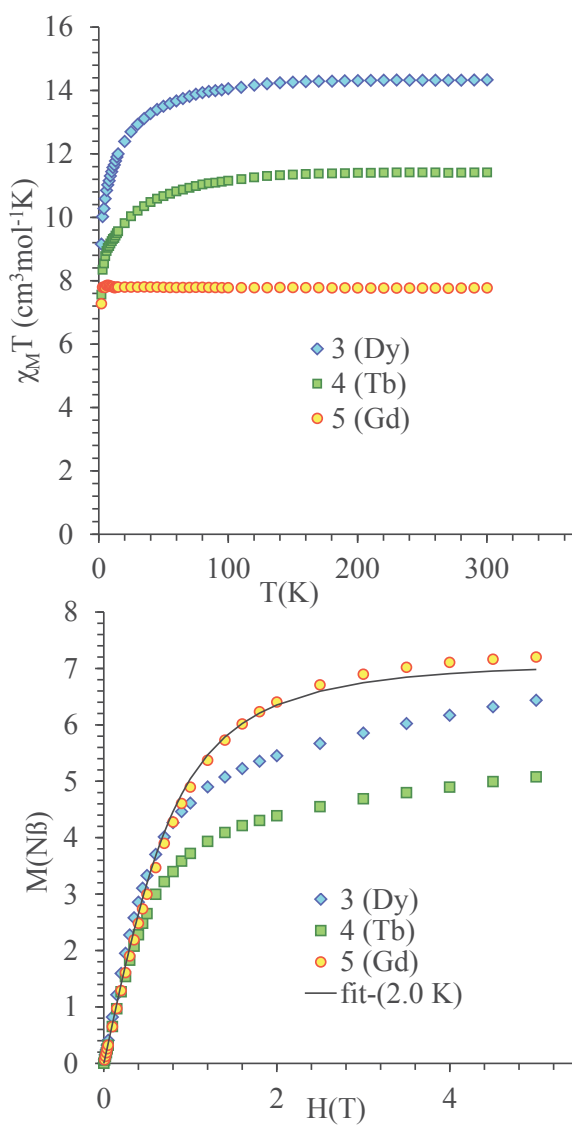

**Figure 5.** top: $\chi_M T$ versus T plots data for **3**, **4** and **5** at 0.1 T dc field. bottom: Magnetization curves for **3–5** at 2 K. The black solid lines correspond to the best fit according to the Brillouin function (S=7/2; g=2.01(3))

## Discussion

The crystal structures of the pentanuclear complexes $(Bu_4N)_5[Ln\{Os(NO)(\mu\text{-}ox)Cl_3\}_4(H_2O)_n]$ (Ln = Y (**2**) and Dy (**3'**) for n = 0, and Ln = Dy (**3**), Tb (**4**) and Gd (**5**) for n = 1] can be described as an onion. In the central part lies the lanthanide atom surrounded by four oxalato ligands in the first shell (Chart 1). Each of them acts as a bridge to the next shell consisting of four osmium atoms: two oxygen atoms of the oxalato ligand coordinate the central lanthanide cation and the two others coordinate one osmium atom. The coordination sphere of the osmium is completed by three chloride anions and a nitrogen atom of the nitrosyl radical (NO). All around this shell lye five tetrabutylammonium cations ($Bu_4N^+$) forming the outer shell. It should be noted that the outer shell is linked via weak Van der Waals interactions to the central $[Ln-Os]^{5-}$ complex. As a consequence, the atomic displacement parameters increase from the lanthanide core to the outer shell. This structure description corresponds to complexes of Y (**2**) and Dy (**3'**) with small radii that are 8-coordinate. For the larger lanthanide ions Tb (**4**) and Gd (**5**), the structure is similar but with one additional molecule of water completing the lanthanide coordination sphere to nine. Interestingly, it was possible to obtain the Dy-Os system with 8-coordination for the Dy (**3'**) lanthanide ion (Table 6). This shows that the radius of the Dy(III) cation is the tilt limit between the coordination numbers 8 and 9 in this lanthanide series. From the crystal structure determination, the average Dy-O distance is 2.372 (0.040) Å for the 8-coordinate system and 2.398 (0.035) Å for the 9-coordinate system. As the Y complex **2** was only observed with the 8-coordination (Y-O = 2.361 (0.040) Å), a radius of approximately 2.38 Å for the central cation seems to be the threshold between the two systems.

**Table 6**: Average bond lengths at 100 K (Å) for **1** (Os), **2** (Y), **3** (Dy), **3'** (Dy), **4** (Tb), and **5** (Gd).

|  | **1** (Os) | **2** (Os-Y) | **3'** (Os-Dy) | **3** (Os-Dy) | **4** (Os-Tb) | **5** (Os-Gd) |
|---|---|---|---|---|---|---|
| Ln-O | / | 2.361(47) | 2.372(40) | 2.394(47) | 2.421(27) | 2.427(25) |
| Ln-O$_{water}$ | / | / | / | 2.40(1) | 2.40(1) | 2.41(1) |
| Os-O | 2.036(18) | 2.079(9) | 2.072(18) | 2.069(86) | 2.077(55) | 2.069(43) |
| Os-N | 1.773(45) | 1.743(8) | 1.745(3) | 1.67(32) | 1.81(11) | 1.59(35) |
| Os-Cl | 2.369(13) | 2.349(21) | 2.350(21) | 2.28(14) | 2.313(93) | 2.31(10) |
| N-O | 1.095(91) | 1.15(1) | 1.154(4) | 1.30(24) | 1.19(19) | 1.19(12) |

The Ln-O distances observed for the oxygen of the water molecule completing the 9-coordination in **3** (Tb), **4** (Dy), and **5** (Gd) is close to the lengths observed for the other Ln-O interactions. The other bond distances are close to those observed for the 8-coordinate cations taking into account the large standard deviations. It should be noted that the Ln-O bond distances are approximately 0.05 Å shorter for Y and Dy (tetragonal) compared to Dy (monoclinic), Tb and Gd. However, the separation Os···Ln is the same for all compared complexes due to the rigidity of the oxalato bridge.

## Conclusion

In the wake of our interest in the magnetism of polynuclear heterometallic complexes and in osmium-nitrosyl based complexes as potential anticancer metal based drugs, we have synthesized a series of osmium-nitrosyl oxalato-bridged lanthanide-centred pentanuclear complexes of the general formula $(Bu_4N)_5[Ln\{Os(NO)(\mu-ox)Cl_3\}_4(H_2O)_n]$ (Ln = Y (**2**) and Dy (**3'**) for n = 0, and Ln = Dy (**3**), Tb (**4**) and Gd (**5**) for n = 1). The crystal structure determination of the series of complexes revealed that the radius of the dysprosium(III) cation is a tilt limit for the central lanthanide. Coordination number eight may be expected for smaller lanthanide ions or yttrium(III) and nine-coordination for larger ones, while both, eight and nine, were observed for dysprosium(III). In these compounds, the $\{Os(NO)\}^6$ moieties are diamagnetic and do not contribute to the magnetism, which is solely due to the central lanthanide ion.

Such complexes may open the way to bifunctional metallapharmaceutics, incorporating all in one a tumoricidal drug with an imaging agent to follow up its migration in the body to the cancerous cell.

## Acknowledgements

This work was done in the frame of an Austrian-French joint project supported in France by ANR (Agence Nationale de la Recherche) through project ANR-09-BLAN-0420-01 (VILYGRu) and in Austria by FWF (Fonds zur Förderung der wissenschaftlichen Forschung) through the project I 374-N19. The PHC Amadeus is also greatly acknowledged for is support. VD thanks the Région Rhône-Alpes for postdoctoral fellowship.

## Dedication

In memoriam of Prof. Jean-Pierre TUCHAGUES

## Supporting Information

CCDC number for CIF file: 1031184 (**1**), 1031157 (**2**), 1031203 (**3'**), 1031242 (**3**), 1031248 (**4**), 1031254 (**5**).